\documentclass[twocolumn]{rbef01A}

\usepackage{cite} 

\usepackage[nomath]{lmodern} 

\usepackage{calrsfs} 
\DeclareMathAlphabet{\pazocal}{OMS}{zplm}{m}{n} 

\usepackage{cancel} 

\titulocabecalho{On the buoyant force in nonuniform gravitational fields}
\autorcabecalho{Nali}

\numeracao{xxxx}
\volume{xx}
\numero{xx}
\ano{20xx}
\doi{http://dx.doi.org/xxxx}
\tipodeartigo{Letter to the Editor}

	\author[1]{Pier Franco Nali}
	\affil[1]{Independent scholar, Via Tempio 29, 09127 Cagliari, Italy.\thanks{Correspondence email address: \href{emailto:pfnali@alice.it}{pfnali@alice.it}}
		} 
	
\titulo{A short note on the buoyant force in nonuniform gravitational fields}

\begin{document}

\begin{primeirapagina}

\begin{center}
\vspace{-12pt}
\small{Received on xxx. Accepted on xxx}
\end{center}

	\begin{abstract}
Lima and Monteiro \cite{LM} derived the buoyant force in a nonuniform gravitational field by applying a gradient version of the divergence theorem to the surface integral of the pressure forces \cite{LI}. Here it is outlined an alternate approach in the framework of fluid mechanics, based on the energy formulations of the equations of state for compressible fluids.
	
	\end{abstract}

	\end{primeirapagina}

\saythanks

In a paper by Lima and Monteiro \cite{LM} it was shown that the validity of the Archimedes’ principle (AP) extends to nonuniform
gravitational fields. Those authors considered an arbitrarily-shaped body fully submerged in a fluid (homogeneous or
stratified) in a nonuniform gravitational field and made use of a gradient version of the divergence theorem applied to the surface integral of the pressure forces \cite{LI}, leading to a volume integral that represents the weight of the displaced fluid.

Locally, the buoyant force field per volume element \(d\,V_f\)\, of the displaced fluid can be expressed as
\begin{equation} \label{eq:eq1}
d\,\mathbf{B}=-\,\nabla p\,\,dV_f=-\,\rho\,\,\mathbf{g}\,\,dV_f=-d\,\mathbf{W}_f\,, 
\end{equation}
where \(p=p\,(\mathbf{r})\) and \(\rho=\rho\,(\mathbf{r})\) are pressure and mass density nonuniform scalar fields and \(\mathbf{g}=\mathbf{g}\,\,(\mathbf{r})\) is a nonuniform gravitational vector field; the buoyant force is compared to the weight per volume element of the displaced fluid \(d\,\mathbf{W}_f\); vectors \(\mathbf{B}\) and \(\mathbf{W}_f\) are in opposite direction to each other. Equation \eqref{eq:eq1} has a corresponding integral form in equation (7) of Lima and Monteiro \cite{LM}. From equation \eqref{eq:eq1} the pressure field plays the role of an effective scalar potential associated with the buoyant force. The body (fully submerged) of weight \(\mathbf{W}_b=\int_{V_f}\rho_b\,\mathbf{g}\,\,d\,V_f\) experiences a net force per volume element
\begin{equation} \label{eq:eq2}
d\,{\mathbf{F}}_b=d\,{\mathbf{W}_b}+d\,{\mathbf{B}}=(\rho_b-\rho)\,\,\mathbf{g}\,\,d\,V_f\,\equiv\rho'\,\,\mathbf{g}\,\,d\,V_f\,,
\end{equation}
where \(\rho_b(\mathbf{r})\) is the mass density of the body and we consider a "deviation" \(\rho'=\rho_b-\rho\) from base state \(\rho\). Accordingly, we call \(p_{F_b}\) the effective scalar potential associated with the net force acting on the body: \(\nabla p_{F_b}=-\,(\rho'/\rho_b)\,\,\mathbf{g}\) (note that the RHS coincides with the relative gravity).

Extending to the product \(\rho'(\mathbf{r})\,\, \mathbf{g\,(\mathbf{r})}\) the integrability condition over \(V_f\) assumed in \cite{LM} for \(\rho\,(\mathbf{r})\,\, \mathbf{g}\,(\mathbf{r})\) (continuous function of \(\mathbf{r}\) in all points of \(V_f\)) we find for the net force experienced by the body
\begin{equation} \label{eq:eq3}
\mathbf{F}_b=\int_{V_f}\,\rho'\,\, \mathbf{g}\,\,d\,V_f=-\int_{V_f}\,\rho_b\nabla p_{F_b}\,\,d\,V_f
\end{equation}
and revert to all the known results proved valid in \cite{LM} in both uniform and nonuniform gravitational fields. In particular, from equation \eqref{eq:eq1} \(\mathbf{B}=-\mathbf{W}_f\), which is the AP as reformulated by Lima and Monteiro (buoyant force "\textit{directed oppositely to the weight of the displaced fluid}"... "\textit{measured at the same place where fluid has been displaced}") \cite{LM}. From equation \eqref{eq:eq2} and equation \eqref{eq:eq3}: 

- at equilibrium (\(\mathbf{F}_b=0\)) \(\mathbf{B}=-\mathbf{W}_b\), so, equation \eqref{eq:eq1} follows from equation \eqref{eq:eq2} as a special case; 

- out of balance (\(\mathbf{F}_b\neq 0\)) the body will sink (for \(\mathbf{F}_b\cdot\mathbf{W}_b>0\)) or float (for \(\mathbf{F}_b\cdot\mathbf{W}_b<0\)). 

Other expected results could also be derived, as Stevin and Pascal laws, and so on.

Lima and Montero also remarked that "\textit{the potential energy minimization technique cannot be used for deriving the AP in these cases since it works only for incompressible fluids}" \cite{LM}. At a closer look this difficulty can be overtaken through a thorough definition of the potential energy (PE) in compressible fluids. In fluid mechanics the equations of state are susceptible of various equivalent formulations, based on internal energy or other thermodynamic potentials, each of which employs a different mix of mechanical and thermal variables. These formulations are suitable for various purposes, including the study of stability conditions. In the formulation based on internal energy, pressure \(p\) or density \(\rho\) are the mechanical variables and entropy \(s\) or temperature \(\varTheta\) the thermodynamic ones. For our purposes it will suffice to consider only the variability of \(p\) or \(\rho\). 

For an ideal fluid in an external gravitational potential \(\varPhi\) the non-kinetic part of the energy density is \(\rho\,(u+\varPhi)\), where \(u=u(s,p)\) is the internal energy per unit mass (see, e.g., \cite{TB}). Exact local expressions for PE density in compressible and incompressible fluids have been obtained since four decades from the very basic principles of fluid mechanics \cite{RT,DA,HI}. As a starting point consider the function
\begin{equation} \label{eq:eq4}
\pazocal{B}(\mathbf{r},s,p)=\varPhi(\mathbf{r})+u(s,p)+\frac{p_0(\mathbf{r})}{\rho(s,p)}\,, 
\end{equation}
that is a rework of the non-kinetic part of the energy density per unit mass. In equation \eqref{eq:eq4} \(p(\mathbf{r})\) is the actual value of the pressure and \(p_0(\mathbf{r})\) the pressure profile function in non-perturbed state. As the PE is always defined relative to a reference, we shall work with the quantity obtained by subtracting from \(\pazocal{B}(\mathbf{r},s,p)\) its value  \(\pazocal{B}(\mathbf{r}_0,s,p(\mathbf{r}_0))\) at a reference point. It is convenient to express this new quantity, denoted by the symbol \(\varPi\), in terms of the enthalpy \(h(s,p)=u(s,p)+p/\rho\) and decompose it in the following two terms:
\begin{equation} \label{eq:eq5}
\varPi_1=h(s,p)-h(s,p_0(\mathbf{r}))-\frac{p-p_0(\mathbf{r})}{\rho}\,, 
\end{equation}
\begin{equation} \label{eq:eq6}
\varPi_2=\varPhi(\mathbf{r})-\varPhi(\mathbf{r}_0)+h(s,p_0(\mathbf{r}))-h(s,p_0(\mathbf{r}_0))\,. 
\end{equation}
\(\varPi_1\) is the "elastic" part of \(\varPi\). It has been called "Available Elastic Energy" (AEE) \cite{RT,DA} and represents the work (of compression/expansion) required to bring an (undisplaced) fluid element from the reference state with pressure \(p_o(\mathbf{r})\) to its actual value \(p\,(\mathbf{r})\) under an adiabatic pressure perturbation \(p'=p-p_0\). \(\varPi_1\) can also be interpreted as an intermediate "reservoir" in the conversion between kinetic energy and \(\varPi_2\), which in turn is the part of PE density available for conversion \cite{RT}. Known as "Available Potential Energy" (APE), the latter represents the work (against buoyant forces) required to move the fluid element from a reference state position to its actual position \cite{RT}. The sum \(\varPi=\varPi_1+\varPi_2\) is the (total) PE density per unit mass.

It should now be observed that the application of the PE minimization technique requires not to consider merely the gravitational potential of the (solid + fluid) system, like, e.g., in \cite{CR}, but to take into account all the components of PE, notably \(\varPi_1\) and \(\varPi_2\), which provide information on the state of the fluid. In order to apply the method properly, the derivatives of \(\varPi_1\) and \(\varPi_2\) have to be calculated separately for the two components as they depend on different independent variables. 

The derivative of \(\varPi_1\), which depends on \(p\), is evaluated by taking the limit for \(p\to p_0\) of the ratio
\[\frac{\varPi_1(p)-\varPi_1(p_0)}{p-p_0}=\frac{h(s,p)-h(s,p_0)}{p-p_0}-\frac{1}{\rho}\frac{\cancel{p-p_0}}{\cancel{p-p_0}}\,,\]
then using the identity \(h_p=1/\rho\) (where the subscript denotes partial derivative), and finally setting the derivative \(d\,\varPi_1/d\, p=0\) to obtain:
\begin{equation} \label{eq:eq7}
\frac{d\varPi_1}{dp}=v(s,p_0)-v(s,p)=0 
\end{equation}
(where \(v=1/\rho\) is the volume per unit mass). From equation \eqref{eq:eq7} we immediately see that the minimum-energy condition for \(\varPi_1\) is only satisfied in incompressible fluids. This result is less significant as it was shown in \cite{DA} that \(\varPi_1=0\) everywhere in the incompressible limit.

For \(\varPi_2\) the independent variables are the coordinates \(x^i\) of the volume element of the fluid, so, in this case, the ratios to consider are of the following form:
\[\frac{\varPi_2(\mathbf{r})-\varPi_2(\mathbf{r}_0)}{x^i-x_0^i}=\frac{\varPhi(\mathbf{r})-\varPhi(\mathbf{r}_0)}{x^i-x_0^i}+\frac{h(s,p_0)-h(s,p_0(\mathbf{r}_0))}{x^i-x_0^i}\,.\]
Setting the derivatives \(\partial\varPi_2\big/\partial x^i=0\) we obtain:
\[\frac{\partial\varPi_2}{\partial x^i}=\frac{\partial\varPhi}{\partial x^i}+\frac{\partial h(s,p)}{\partial p}\,\cdot\,\frac{\partial p}{\partial x^i}=\frac{d\varPhi}{d x^i}+\frac{1}{\rho}\,\cdot\,\frac{\partial p}{\partial x^i}=0\]
or, in vector form,
\begin{equation} \label{eq:eq8}
\nabla{\varPi_2}=\nabla{\varPhi}+\frac{1}{\rho_0}\,\cdot\,\nabla{p_0}=0\,,
\end{equation}
that is the hydrostatic equilibrium equation, equivalent to equation \eqref{eq:eq1} (remembering that \(\nabla{\varPhi}=-\mathbf{g}\,\)). The 0 subscript indicates that the reference state is hydrostatic equilibrium. The condition expressed by equation \eqref{eq:eq8} can be introduced \textit{a priori} to obtain the potentials, as in \cite {DA}, or, on the reverse, obtained from the potentials as we did.

The energy formulation in terms of potentials has further advantages: the positive-definiteness of both \(\varPi_1\) and \(\varPi_2\) \cite{RT,DA}, and their convexity as functions of their explicit variables for compressible fluids \cite{DA}. The convexity corresponds to the condition for static stability to small perturbations (see, e.g., \cite{HH}). 

The case (solid+fluid) can be traced back to the previous argument with a simple trick. As the hydrostatic balance is assumed as reference state, imagine the fully submerged solid (bounded by a rigid impermeable surface) held in place by an external force \(-\mathbf{F}_b\) opposite to the unbalanced resulting net force \(\mathbf{F}_b\). Under a pressure disturbance, equation \eqref{eq:eq7} still holds locally for \(\varPi_1\) in the case of rigid body, thought of as a special limiting case of fluid. As no energy change is involved, we put this case aside. 

As for \(\varPi_2\), equation \eqref{eq:eq8} continue to hold locally in the form
\begin{equation} \label{eq:eq9}
\rho_b\,\nabla{\varPhi_e}+\nabla{p_b}=0\,,
\end{equation}
if we substitute an effective potential \(\varPhi_e=\varPhi-p_{F_b}\) in place of \(\varPhi\), assuming the counterbalancing force \(-\mathbf{F}_b\) to be a conservative force of the same type of weight \(\mathbf{W}_b\), namely \(-\int_{V_f}\rho_b\nabla{\varPhi_e} d\,V_f=-\mathbf{F}_b+\mathbf{W}_b\) and \(-\int_{V_f}\rho_b\nabla{\varPhi}\,\,d\,V_f=\mathbf{W}_b\,.\) 
It is also assumed that the object and the environment experience the same pressure, so that \(\nabla{p_b}=\nabla{p}\) (condition that is called "rapid adjustment" between external and internal pressure). Then, exploiting equation \eqref{eq:eq8}, equation \eqref{eq:eq9} can be rearranged to give
\begin{equation} \label{eq:eq10}
(\rho-\rho_b)\,\nabla{\varPhi}=(\rho_b-\rho)\,\,\mathbf{g}=\rho'\,\mathbf{g}=\frac{1}{V_f}\mathbf{F}_b\,,
\end{equation}
that is equivalent to equation \eqref{eq:eq3} if \(\mathbf{F}_b\) is a constant force. Note that equation \eqref{eq:eq9} or \eqref{eq:eq10} is also valid when the body is thought to enclose within its surface a  fluid other than the environment it is immersed in. In this case the body exhibits a variety of behaviors depending on the density lapse rates of the two fluids, enclosed and external. As the densities are functions of \(\mathbf{r}\) the level of neutral buoyancy (LNB) is obtained as solution of the equation \(\rho'(\mathbf{r})=0\) for some \(\mathbf{r}\), if any. The stability to vertical displacements of a fluid element in a bulk can be analyzed quite easily, even without requiring mathematics (see, e.g., \cite{BL}). As a rule of thumb, if the bulk has the higher density lapse rate, it exhibits a convex effective potential towards the immersed fluid element, which is accordingly vertically stable; vice versa, with inverted lapse rates, the effective potential looks concave and the current vertical position attained by the fluid element either out of balance or unstable.

We have an example of vertical stability considering a balloon with semi-rigid casing ascending in a column of highly compressible air. Should the balloon (assumed of invariable volume) reach a LNB, climbing further it encounters less dense air. As a result, the upthrust weakens and the balloon is eventually brought down by its own weight. In the descent, the air becomes denser and denser until the upthrust resumes. The cycle repeats and the balloon experiences a series of vertical oscillations around the LNB.

A classical example of vertical instability is demonstrated by the toy known as Cartesian diver. The air trapped in the diver (highly compressible) react to the pressure variations of the surrounding water (which is assumed incompressible) changing in volume for the same mass. If the diver rises the pressure acting on it decreases and the the trapped air will expand. As a result, the diver becomes positively buoyant, rising more and more quickly. Conversely, if the diver drops the pressure it feels increases and the air bubble shrinks; the diver becomes less buoyant and the drop will accelerate more and more. In both cases a departure from equilibrium is experienced by the diver.

More formally, from eq. \eqref{eq:eq10} the acceleration of the fully submerged body due to the effect of the scalar potential \(p_{F_b}\) associated to the unbalanced force is expressed by \[\frac{\rho'}{\rho_b}\mathbf{g}=-\nabla{p_{F_b}}=\frac{\mathbf{F}_b}{\rho_b V_f}.\]
This allows you to check immediately that the method we have followed yields the correct buoyant force depending on the relative gravity of the body on the LHS.

For vertical displacements the LNB(s) solution(s) of \(\rho_b(z)=\rho(z)\) corresponds (correspond) to the extrema of the effective potential \(p_{F_b}\), calculated from
\[\frac{\partial p_{F_b}}{\partial z}=\frac{\rho'}{\rho_b}g_z=\rho' v_b\,g_z=0\,.\]
Said \(z_0\) an extremum corresponding to one LNB solution, it is a minimum or a maximum of the effective potential depending on the sign of the second derivative
\[\frac{\partial^2 p_{F_b}}{\partial z^2}=\frac{\partial}{\partial z}\left(\rho' v_b\,g_z\right).\]
Assuming \(g_z\simeq g_z(z_0)\) around \(z_0\) and  \(v_b\simeq v(z_0)+\\(\partial v_b/\partial z)_{z_0}\Delta z\)
the problem reduces to the determination of the sign of the relative lapse rate \((\partial\rho'/\partial z)_{z_0}\). Remembering that \(\rho'=\rho_b-\rho\), in the case of a solid object (\(\rho_b\simeq \text{const}\)), like a balloon, vertically displacing on an atmosphere with air density decreasing with the altitude (\((\partial\rho/\partial z)<0)\) the density lapse rate of the air bulk largely prevails. Then \((\partial\rho'/\partial z)\simeq -(\partial\rho/\partial z)>0\), the effective potential exhibits a minimum (it looks like a convex function) and the object attains vertical stability. It is straightforward to show that in the proximity of \(z_0\) the effective potential \(p_{F_b}\) is a quadratic function of the displacement \(z-z_0\). This accounts for the LNB oscillations of the body.

Conversely, for a bubble of a compressible fluid vertically displacing on a liquid column (case of the Cartesian diver) the density lapse rate of the bulk is quite negligible, so that \((\partial\rho'/\partial z\simeq\partial\rho_b/\partial z)<0\). The extemum of the effective potential is now a maximum (the potential looks like a concave function), indicating that at \(z=z_0\) the equilibrium position is unstable.

It may seem surprising that such a long-standing and well-established physical principle such as the AP - arguably the oldest physical law - is still a subject of current investigations. Its validity has been widely confirmed at the mesoscopic and macroscopic scales. Recently \cite{JP} the local AP has been derived microscopically, in the framework of statistical mechanics, confirming the generalized form of the principle worked out by Lima and Montero \cite{LM}, valid for any non-uniform pressure field. On the other hand, has been shown \cite{RP} that AP suffers limitations in the nanoworld.

\section*{Acknowledgements}
I would like to thank the reviewers for their thoughtful suggestions.



\end{document}